\documentclass[notitlepage,prd,showpacs,superscriptaddress,nofootinbib,floatfix,showkeys,10pt]{revtex4-1}
\usepackage{graphicx}
\usepackage{amsmath}
\usepackage{bm}
\usepackage{yhmath}
\usepackage{mathtools}
\usepackage{wasysym}
\usepackage{subfigure}
\usepackage{xcolor}
\definecolor{CherryRed}{rgb}{.65,0,.2}
\definecolor{RubyRed}{rgb}{.88,0.07,.3}
\definecolor{CralRed}{rgb}{1,0.25,.25}
\definecolor{CobaltBlue}{rgb}{0,0.28,.67}
\definecolor{RoyalBlue}{rgb}{0.25,0.41,.88}
\definecolor{EmeraldGreen}{rgb}{0.31,0.78,.47}
\definecolor{EmeraldGreen}{rgb}{0.31,0.78,.47}
\definecolor{LimeGreen}{rgb}{50,205,50}
\definecolor{ForestGreen}{RGB}{34,139,34}
\definecolor{PineGreen}{RGB}{1,121,111}\usepackage{cases}
\usepackage[colorlinks=true, linkcolor=EmeraldGreen,citecolor=RoyalBlue,urlcolor=ForestGreen,linkcolor=ForestGreen]{hyperref}
\usepackage{subfigure}
\usepackage{times}
\usepackage{booktabs,bm}
\usepackage{slashed}
\usepackage{amsfonts,amssymb,stmaryrd,latexsym,amsmath}
\usepackage{textcomp}
\usepackage{multirow}
\usepackage{cancel}
\usepackage{array}
\usepackage{hyperref}
\usepackage{pgf}
\usepackage{orcidlink}

\def\SU3{{\text{SU(3)}_{\rm F}}}

\def \pcs4338{{P_{\psi s}^\Lambda(4338)^0}}

\begin{document}
	
	\title{\textcolor{CobaltBlue}{    Examining  possible doubly topped baryon configurations}}

	\author{M.~Shekari Tousi\orcidlink{0009-0007-7195-0838}}
	\email{ marzie.sh.tousi@ut.ac.ir }
	\affiliation{Department of Physics, University of Tehran, North Karegar Avenue, Tehran 14395-547,  Iran }

	\author{K.~ Azizi\orcidlink{0000-0003-3741-2167}}
	\email{kazem.azizi@ut.ac.ir}
	\thanks{Corresponding author}
	\affiliation{Department of Physics, University of Tehran, North Karegar Avenue, Tehran 14395-547,  Iran }
	\affiliation{Department of Physics,  Dogus University, Dudullu-\"{U}mraniye, 34775 Istanbul, T\"urkiye }

	\date{\today}

	\begin{abstract}

We present a comprehensive theoretical assessment of  the masses of possible baryonic configurations characterized by the presence of two heavy top quarks,  including $\Xi_{ttu}$, $\Xi_{ttd}$, $\Omega_{tts}$, $\Omega_{ttc}$, and $\Omega_{ttb}$ systems. This analysis is rigorously executed within the specialized framework of  two-point $\mathrm{QCD}$ sum rules, focusing on their predicted ground state masses. Our interest in these systems arises from recent CMS and ATLAS reports indicating a pseudoscalar excess close to $t\bar{t}$ threshold. Our evaluation incorporates both perturbative terms and nonperturbative effects, including condensate contributions up to dimension eight.  Based on our results, the extracted central masses for all channels are slightly above the sum of the constituent quark masses, which is consistent with the inherent uncertainties of the method. These quantitative predictions provide a useful first-principle  theoretical reference, which may help  future experimental searches for such heavy configurations at the LHC and inform sensitivity studies at next-generation facilities such as the FCC.
\end{abstract}

	\maketitle
	\renewcommand{\thefootnote}{\#\arabic{footnote}}
	\setcounter{footnote}{0}


\section {Introduction}\label{sec:one}
As the most massive particle in the Standard Model (SM),  the top quark plays a central role in modern high-energy physics. Since its first experimental identification in the mid-1990s \cite{CDF:1995wbb,D0:1995jca}, it has stood out not only for its unusually large mass scale but also for the distinctive way it interacts within the strong and electroweak sectors. These characteristics make it a key probe for exploring the fundamental behavior of matter at the highest energies. The importance of the top quark extends far beyond its role as a stringent probe of the SM. Its properties also open a direct window onto potential physics beyond the SM. 
Among all quarks, the top quark is particularly remarkable for its extremely rapid decay, on the order of $  10^{-25}$ s \cite{Bigi:1986jk,Bernreuther:2008ju}. Indeed, the top quark is characterized by a relatively large decay width, $\Gamma_t \simeq 1.4$ GeV, and therefore systems containing top quarks are naturally expected to possess sizable widths due to the large width of their constituent top quarks. This unusual combination of an exceptionally large mass and a vanishingly short lifetime has positioned the top quark at the center of both SM and BSM investigations for many years \cite{Schrempp:1996fb,Bhat:1998cd,Cvetic:1997eb,Atwood:2000tu,Plehn:2011tg,Chakraborty:2003iw,Merkel:2004td,Wagner:2005jh,Quadt:2006dqn,Kats:2009bv,Galtieri:2011yd,Palle:2012mr,Schilling:2012dx,Hagiwara:2016rdv}.

New measurements have   challenged the long-held view that the top quark cannot participate in a bound state formation. Notably, the CMS and ATLAS experiments at the LHC have observed an unexpected enhancement in the vicinity of the  production threshold, compatible with the appearance of a pseudoscalar, quasi-bound toponium like configuration \cite{CMS:2025kzt,ATLAS:2025mvr}. One of the distinguishing quantum features of $t\bar{t}$   produced near threshold is its pronounced spin correlation. This behavior, observable even at very high energies, may signal the presence of near-bound dynamics or reflect substantial quantum entanglement within the system. 
Although the entanglement of the  pair cannot be measured directly, it can still leave observable imprints on the system’s behavior near threshold, an effect made especially relevant by the extremely short lifetimes of the top and antitop quarks. The recent experimental excess, established through decay channel analyses with a significance surpassing $5\sigma$ \cite{CMS:2025kzt}, quickly sparked extensive theoretical activity across the community \cite{Fuks:2025toq,Matsuoka:2025jgm,Sjostrand:2025qez,Fuks:2025wtq,Goncalves:2025hyx,Zhu:2025ezg,Zhang:2025fdp,Lopez:2025kog,Thompson:2025cgp,Zhang:2025xxd,Shao:2025dzw,Bai:2025buy,Ellis:2025nkm,LeYaouanc:2025mpk,Xiong:2025iwg,Fu:2025yft,Fu:2025zxb,Fuks:2025sxu,Afik:2025ejh,Luo:2025psq}. An important point is that many earlier studies had already proposed the possibility of a toponium state and explored its expected characteristics \cite{Fadin:1987wz,Kuhn:1987ty,Barger:1987xg,Strassler:1990nw,Fadin:1990wx,Fadin:1994pj,Hoang:2000yr,Hagiwara:2008df,Penin:2005eu,Sumino:2010bv,Kiyo:2008bv,Beneke:2015kwa,Kawabata:2016aya,Reuter:2018rbq,Fuks:2021xje,Aguilar-Saavedra:2024mnm,Jiang:2024fyw,Jafari:2025rmm,Francener:2025tor,dEnterria:2025ecx,Garzelli:2024uhe,Wang:2024hzd,Fuks:2024yjj}.

Beyond that, the quark model provides a systematic framework for anticipating the spectrum of baryons containing one, two, or three heavy quarks \cite{GellMann:1964nj}. States with a single heavy quark have long been established experimentally, but systems composed of two heavy quarks remained elusive for many years, creating a notable gap between theoretical expectations and observations. This situation changed in 2017, when the LHCb collaboration identified the doubly charmed baryon $ \Xi^{++}_{cc} (3621)$ through the $ \Xi^{++}_{cc}\rightarrow \Lambda^+_{c} K^- \pi^+ \pi^+$ decay mode \cite{LHCb:2017iph}. A second measurement in 2018, using the channel $ \Xi^{++}_{cc}\rightarrow \Xi^+_{c} \pi^+$,  provided further confirmation of its existence \cite{LHCb:2018pcs}. These findings spurred extensive theoretical efforts employing a wide range of approaches to investigate the characteristics such as masses, decay properties, and internal dynamics of baryons containing two heavy quarks \cite{ShekariTousi:2024mso,ShekariTousi:2025fjf,Tousi:2024usi,ShekariTousi:2025xox, Ebert:2002ig,Zhang:2008rt,Wang:2010hs,Lu:2017meb,Rahmani:2020pol,Yao:2018ifh,Aliyev:2022rrf,Aliev:2012iv,Aliev:2019lvd,Aliev:2012ru,Padmanath:2019ybu,Brown:2014ena,Giannuzzi:2009gh,Shah:2017liu,Shah:2016vmd,Yoshida:2015tia,Li:2022ywz,Wang:2010it,Ortiz-Pacheco:2023kjn,Wang:2018lhz,Bagan:1992za,Alrebdi:2022lat,Wang:2010vn}. In contrast, baryons built entirely from heavy quarks have not yet been detected, and their eventual discovery remains an important objective for upcoming experimental programs.

Recent studies on the possible formation of toponium have revived interest in whether the SM could also accommodate a baryonic state built from two top quarks. Such a particle, if it can form, would surpass all known baryons in mass and offer an exceptional setting for examining the QCD dynamics at energy scales far beyond those of conventional hadrons. While the top quark’s very rapid decay (characterized by a width of approximately 1.3 GeV) has long been viewed as preventing it from participating in hadron formation, some theoretical investigations suggest that the characteristic formation time  of  a triply top baryon   could be compatible with the brief lifetime of the top quark  \cite{Fu:2025yft,Fu:2025zxb,Xiong:2025iwg,Jia:2006gw,Najjar:2025bby}. This opens the possibility that a   system with two top quarks might still arise before the individual constituents decay.

In this study, we investigate the masses   of   baryons containing two heavy top quarks, focusing on the $\Xi_{ttu}$, $\Xi_{ttd}$, $\Omega_{tts}$, $\Omega_{ttc}$ and $\Omega_{ttb}$  baryons. Our analysis is carried out utilizing the QCD sum rule formalism, a method derived directly from the QCD Lagrangian that has proven reliable in describing the properties of systems containing heavy quarks. Previous applications of this framework have shown strong consistency with measured characteristics of heavy hadrons \cite{Aliev:2009jt,Aliev:2010uy,Aliev:2012ru,Agaev:2016mjb,Azizi:2016dhy}.
To determine the masses of the states considered, we construct appropriate two-point correlation functions and evaluate them through the operator product expansion (OPE), incorporating nonperturbative effects up to dimension eight condensates. The sum rule procedure further includes Borel transformation and a continuum subtraction scheme informed by quark–hadron duality. Together, these elements allow for a controlled extraction of the masses of the doubly topped baryons under study. The findings presented here aim to support the theoretical interpretation of upcoming measurements at the LHC and to guide analyses at future experimental facilities. 

The paper is organized as follows. Section  \ref{sec:two}  outlines the methodology based on QCD sum rules and summarizes the computational setup. Section  \ref{sec:three}  presents the mass predictions obtained for doubly topped baryons and provides a comparative discussion. Section  \ref{sec:four}  offers the concluding remarks and an overview of the implications of our study.  

\section {MASS CALCULATION WITHIN THE QCD SUM RULES}\label{sec:two}

In this work, we use the QCD sum rule method to evaluate the masses of the doubly topped   spin-1/2  baryons, including $\Xi_{ttu}$, $\Xi_{ttd}$, $\Omega_{tts}$, $\Omega_{ttc}$ and $\Omega_{ttb}$ baryons. The sum rule formalism offers a consistent way to link measurable hadronic properties to the underlying quark–gluon dynamics described by QCD. Our analysis proceeds in two complementary steps. On the phenomenological side, the relevant correlation function is expressed in terms of hadronic parameters such as masses and decay constants. On the theoretical side, the same correlator is expanded using the OPE, where perturbative contributions and gluonic vacuum effects are systematically included. Equating these two representations through quark–hadron duality and a dispersion relation yields the sum rules from which the physical quantities can be derived. The final extraction of the mass and related parameters follows from matching the coefficients associated with the appropriate Lorentz structures in both descriptions.
To establish the QCD sum rules for the doubly    topped baryons, one must first specify a suitable correlation function. The analysis begins by introducing correlators built from time ordered products of the relevant interpolating currents. These objects serve as the foundation of the sum rule method, providing the link between hadronic observables and the underlying quark–gluon dynamics  defined in the form:

\begin{eqnarray}
	\Pi (q)&=&i\int d^{4}xe^{iqx}\langle 0|\mathcal{T}\{\eta (x)\bar{\eta} (0)\}|0\rangle. 
		 \label{eq:CF1}
	\end{eqnarray}
	In this expression, the symbol $\eta(x)$ denotes the interpolating current associated with the doubly topped baryon states. The operator $\mathcal{T}$specifies time ordering, and q refers to the four momentum carried by the hadronic state. Establishing the QCD sum rules for the systems considered requires introducing suitable interpolating currents for each channel. The most general form of the interpolating currents of    symmetric  (concerning the heavy quark exchange)   doubly heavy baryons is 
\begin{align}\label{eq:CorrF2}
	\eta (x)=\frac{1}{\sqrt{2}} \epsilon_{abc} \Bigg\{(Q^{a^T} C q^b)\gamma_5 Q'^c  +(Q'^{a^T} C q^b)\gamma_5 Q^c +\beta (Q^{a^T} C \gamma_5 q^b) Q'^c+ \beta  (Q'^{a^T} C \gamma_5 q^b) Q^c\Bigg\}.
\end{align}
	In Eq.  (\ref{eq:CorrF2}), for the $\Xi_{ttu}$, $\Xi_{ttd}$, and $\Omega_{tts}$ baryons, the quark fields $Q$ and $Q'$ are identical and are both taken to represent the 
	top quark field   and $q$ represents the light quark fields. $C$  stands for the charge conjugation matrix, while $a$, $b$ and $c$ label the color components of the quark fields.  Here, $\beta$
	represents a   mixing parameter that must be determined within the analysis; setting $\beta =-1$	yields the form known as the Ioffe current \cite{Ioffe:1981kw}. The   interpolating
	current for the triply heavy $\Omega_{ttc}$ and $\Omega_{ttb}$ baryons is  constructed as follows:
	
	\begin{equation}\label{eq:CorrF3}
		\eta (x)=2\varepsilon^{abc}\Bigg\{\Big(Q^{aT}(x)CQ^{'b}(x)\Big)\gamma_{5}Q^c(x)+
		\beta\Big(Q^{aT}(x)C\gamma_{5}Q^{'b}(x)\Big)Q^c (x)\Bigg\}.
	\end{equation}
	In this current definition, $Q$	illustrates the  top quark field,  $Q'$ indicates bottom or charm quark. 
	
	 To proceed, the two-point correlation function built from these interpolating currents is evaluated in two complementary frameworks: the phenomenological representation, which encodes the hadronic dynamics, and the QCD side description based on the OPE.

 \subsection{Phenomenological representation } 	
 In the sum rule approach, the phenomenological representation of the correlator is obtained by inserting a full set of hadronic states whose quantum numbers coincide with those of the chosen interpolating current. After carrying out the spacetime integration and separating the lowest-lying state from higher resonances and continuum contributions, the correlation function associated with the considered baryonic states,  takes the following hadronic form:
 
\begin{align}
	\Pi ^{\mathrm{Phys}}(q)=\frac{\langle0|\eta |B(q,s)\rangle\langle B(q,s)|\bar{\eta }|0\rangle}{m_B^2-q^2}
 +\cdots.
	\label{Eq:cor:Phys}
	\end{align}
	The symbol $|B (q,s)\rangle$ denotes the single–particle ground state of the baryon, while the terms represented by $\mathbf{\cdots}$  account for contributions originating from excited resonances and the continuum sector. The relevant matrix elements appearing in Eq.~(\ref{Eq:cor:Phys}) can be introduced in the following manner:
	\begin{eqnarray}
		\langle 0|\eta|B (q,s)\rangle&=&\lambda u(q,s).
	\end{eqnarray}  
	In this expression, $\lambda$ designates the coupling strength of the baryonic state, and $u(q,s)$ denotes the corresponding Dirac spinor. This matrix element enters Eq.~(\ref{Eq:cor:Phys}), where the standard spin summation for Dirac spinors is
	\begin{eqnarray}
		\sum_{s}u(q,s)\bar{u}(q,s)=(\not\!q+m_B),
	\end{eqnarray}  
	and the correlation function becomes
	\begin{eqnarray}
		\Pi^{\mathrm{Phys}}(q)=\frac{\lambda^2(\not\!q+m_B)}{m_B^2-q^2} +\cdots.
		\label{Eq:cor:Phys1}
	\end{eqnarray}
	 
	Applying the Borel transformation to the phenomenological expression yields the final form of the physical contribution,
 \begin{eqnarray}
 	\tilde{\Pi}^{\mathrm{Phys}}(q)=\lambda^2(\not\!q+m_B)e^{-\frac{m_B^2}{M^2}}+\cdots,
 	\label{Eq:cor:Fin}
 \end{eqnarray}
	where $\tilde{\Pi}^{\mathrm{Phys}}(q)$  represents the Borel transformed correlation function.
	
	   \subsection{OPE representation}	
	   To evaluate the QCD contribution, we start from correlation function Eq. (\ref{eq:CF1}) by substituting the interpolating currents of considered baryons, Eqs.~(\ref{eq:CorrF2}) and (\ref{eq:CorrF3}), into this function. The subsequent steps require performing all admissible contractions of the quark fields according to Wick’s theorem. After carrying out these contractions, the resulting quark operators are replaced by the corresponding heavy and light quark propagators in coordinate space, whose explicit representations are given below 
	   
	   \begin{eqnarray}
	   	&&S_{q}^{ab}(x)=i\delta _{ab}\frac{\slashed x}{2\pi ^{2}x^{4}}-\delta _{ab}%
	   	\frac{m_{q}}{4\pi ^{2}x^{2}}-\delta _{ab}\frac{\langle \overline{q}q\rangle
	   	}{12}+i\delta _{ab}\frac{\slashed xm_{q}\langle \overline{q}q\rangle }{48}%
	   	-\delta _{ab}\frac{x^{2}}{192}\langle \overline{q}g_{s}\sigma Gq\rangle
	   	\notag \\
	   	&&+i\delta _{ab}\frac{x^{2}\slashed xm_{q}}{1152}\langle \overline{q}%
	   	g_{s}\sigma Gq\rangle -i\frac{g_{s}G_{ab}^{\alpha \beta }}{32\pi ^{2}x^{2}}%
	   	\left[ \slashed x{\sigma _{\alpha \beta }+\sigma _{\alpha \beta }}\slashed x%
	   	\right] -i\delta _{ab}\frac{x^{2}\slashed xg_{s}^{2}\langle \overline{q}%
	   		q\rangle ^{2}}{7776}  \notag \\
	   	&&-\delta _{ab}\frac{x^{4}\langle \overline{q}q\rangle \langle
	   		g^2_{s}G^{2}\rangle }{27648}+\cdots ,  \label{eq:A1}
	   \end{eqnarray}%
	   and
	   \begin{eqnarray}
	   	&&S_{Q}^{ab}(x)=i\int \frac{d^{4}k}{(2\pi )^{4}}e^{-ikx}\Bigg \{\frac{\delta
	   		_{ab}\left( {\slashed k}+m_{Q}\right) }{k^{2}-m_{Q}^{2}}-\frac{%
	   		g_{s}G_{ab}^{\alpha \beta }}{4}\frac{\sigma _{\alpha \beta }\left( {\slashed %
	   			k}+m_{Q}\right) +\left( {\slashed k}+m_{Q}\right) \sigma _{\alpha \beta }}{%
	   		(k^{2}-m_{Q}^{2})^{2}}  \notag  \label{eq:A2} \\
	   	&&+\frac{g_{s}^{2}G^{2}}{12}\delta _{ab}m_{Q}\frac{k^{2}+m_{Q}{\slashed k}}{%
	   		(k^{2}-m_{Q}^{2})^{4}}+\frac{g_{s}^{3}G^{3}}{48}\delta _{ab}\frac{\left( {%
	   			\slashed k}+m_{Q}\right) }{(k^{2}-m_{Q}^{2})^{6}}\left[ {\slashed k}\left(
	   	k^{2}-3m_{Q}^{2}\right) +2m_{Q}\left( 2k^{2}-m_{Q}^{2}\right) \right] \left(
	   	{\slashed k}+m_{Q}\right) +\cdots \Bigg \}. \notag \\
	   	&&
	   \end{eqnarray}
Here, $G_{\mu\nu}$ denotes the gluonic field strength tensor, and its color components are written as $G_{ab}^{\alpha\beta}=G_A^{\alpha\beta}t^A_{ab}$ with $t^A=\lambda^A/2$ and we define the gluonic terms:  $G^{2}=G_{\alpha \beta }^{A}G_{A}^{\alpha \beta }$ and $G^{3}=f^{ABC}G_{\alpha
	\beta }^{A}G^{B\beta \delta }G_{\delta }^{C\alpha }$. The matrices $\lambda
	^{A}$ are the Gell-Mann generators of color $SU_{c}(3)$ group and $f^{ABC}$  represent its antisymmetric structure constants, with indices A, B, C= 1,2, ..., 8.
As a result, the  following form for the doubly top $\Xi_{ttu}$, $\Xi_{ttd}$ and $\Omega_{tts}$ baryons in terms of the heavy and light propagators is obtained :
	   \begin{align} \label{shekari}
	   	\Pi (q)&=i   \epsilon_{a b c} \epsilon_{a^{\prime} b^{\prime} c^{\prime}} \int d^{4} x e^{i q x} \Bigg\{-\gamma_{5} S_{Q}^{c b^{\prime}} S_{q}^{\prime b a^{\prime}} S_{Q}^{a c^{\prime}} \gamma_{5}-\gamma_{5} S_{Q }^{c b^{\prime}} S_{q}^{\prime b a^{\prime}} S_{Q}^{a c^{\prime}} \gamma_{5} +\gamma_{5} S_{Q }^{c c^{\prime}} \gamma_{5} Tr\Big[S_{Q}^{a b^{\prime}} S_{q}^{\prime b a^{\prime}}\Big]\nonumber\\
	   	&+\gamma_{5} S_{Q}^{c c^{\prime}} \gamma_{5} Tr\Big[S_{Q }^{a b^{\prime}} S_{q}^{\prime b a^{\prime}}\Big] +\beta \Bigg(-\gamma_{5} S_{Q}^{c b^{\prime}} \gamma_{5} S_{q}^{\prime b a^{\prime}} S_{Q }^{a c^{\prime}}-\gamma_{5} S_{Q }^{c b^{\prime}} \gamma_{5} S_{q}^{\prime b a^{\prime}} S_{Q}^{a c^{\prime}}-S_{Q}^{c b^{\prime}} S_{q}^{\prime b a^{\prime}} \gamma_{5} S_{Q }^{a c^{\prime}} \gamma_{5}\nonumber\\
	   	&-S_{Q }^{c b^{\prime}} S_{q}^{\prime b a^{\prime}} \gamma_{5} S_{Q}^{a c^{\prime}} \gamma_{5} +\gamma_{5} S_{Q}^{c c^{\prime}}Tr\Big[S_{Q}^{a b^{\prime}} \gamma_{5} S_{q}^{\prime b a^{\prime}}\Big]+S_{Q }^{c c^{\prime}} \gamma_{5} Tr\Big[S_{Q}^{a b^{\prime}} S_{q}^{\prime b a^{\prime}} \gamma_{5}\Big]+\gamma_{5} S_{Q}^{c c^{\prime}} Tr\Big[S_{Q }^{a b^{\prime}} \gamma_{5} S_{q}^{\prime b a^{\prime}}\Big]\nonumber\\
	   	&+S_{Q}^{c c^{\prime}} \gamma_{5} Tr\Big[S_{Q }^{a b^{\prime}} S_{q}^{\prime b a^{\prime}} \gamma_{5}\Big]\Bigg) +\beta^{2}\Bigg(-S_{Q}^{c b^{\prime}} \gamma_{5} S_{q}^{\prime b a^{\prime}} \gamma_{5} S_{Q }^{a c^{\prime}}-S_{Q }^{c b^{\prime}} \gamma_{5} S_{q}^{\prime b a^{\prime}} \gamma_{5} S_{Q}^{a c^{\prime}}+S_{Q }^{c c^{\prime}} Tr\Big[S_{q}^{b a^{\prime}} \gamma_{5} S_{Q}^{\prime a b^{\prime}} \gamma_{5}\Big]\nonumber\\
	   	&+S_{Q}^{c c^{\prime}} Tr\Big[S_{q}^{b a^{\prime}} \gamma_{5} S_{Q }^{\prime a b^{\prime}} \gamma_{5}\Big]\Bigg)\Bigg\}.
	   \end{align}
For the $\Omega_{ttc}$ and $\Omega_{ttb}$ baryons,  we have:	   
	   	\begin{align}\label{eqQCD1}
	   	\Pi(q)&=4i\ \epsilon_{a b c} \epsilon_{a^{\prime} b^{\prime} c^{\prime}} \int d^4x e^{i q x} \Bigg\{-\gamma_{5}
	   	S^{c'b}_{Q}S'^{b'a}_{Q'} S^{a'c}_{Q} \gamma_{5}+
	   	\gamma_{5}S^{c'c}_{Q} \gamma_{5}Tr\Big[S^{a'b}_{Q} S'^{b'a}_{Q'} \Big]
	   	\nonumber\\
	   	&+ \beta\Bigg( -\gamma_{5}S^{c'b}_{Q} \gamma_{5}S'^{b'a}_{Q'} S^{a'c}_{Q} 
	   	-S^{c'b}_{Q} S'^{b'a}_{Q'} \gamma_{5}S^{a'c}_{Q} \gamma_{5}
	   	+\gamma_{5}S^{c'c}_{Q} Tr\Big[S^{a'b}_{Q} \gamma_{5}S'^{b'a}_{Q'} \Big]\nonumber\\
	   	&+ S^{c'c}_{Q} 
	   	\gamma_{5}Tr\Big[S^{a'b}_{Q} S'^{b'a}_{Q'} \gamma_{5}\Big]\Bigg)+
	   	\beta^2\Bigg( -S^{c'b}_{Q} \gamma_{5}S'^{b'a}_{Q'} \gamma_{5}S^{a'c}_{Q} +S^{c'c}_{Q} 
	   	Tr\Big[S^{b'a}_{Q'} \gamma_{5}S'^{a'b}_{Q} \gamma_{5}\Big]
	   	\Bigg)
	   	\Bigg\},
	   \end{align}
	   where $S' = CS^T C$.
	   To derive the final QCD representation of the sum rule, the analysis begins with the insertion of the relevant propagator expressions, after which the correlation function is evaluated through successive Fourier and Borel transformations. The contribution of higher excited states and the continuum is then removed by invoking quark–hadron duality. The resulting expression for the QCD side is ultimately written in the following form for the   $\Xi_{ttu}$, $\Xi_{ttd}$ and $\Omega_{tts}$ baryons:
	   
	   	  \begin{equation}\label{eqQCD4}
	   	\tilde\Pi_{1(2)}(q)=\int_{4m^2_{Q}}^{s_0} ds e^{-s/M^2} \rho_{1(2)} (s) + \Gamma _{1(2)} (M^2),
	   \end{equation}
and  for the $\Omega_{ttc}$ and $\Omega_{ttb}$ baryons we  have	   	   
	  \begin{equation}\label{eqQCD5}
	  	\tilde\Pi'_{1(2)}(q)=\int_{(2m_{Q}+m_{Q\prime})^2}^{s_0} ds e^{-s/M^2} \rho'_{1(2)} (s) + \Gamma' _{1(2)} (M^2).
	  \end{equation}
	  In the above formulation, the labels $1$ and $2$ correspond to the invariant amplitudes multiplying the Lorentz structures $\not q$ and the identity operator $I$, respectively. The quantity $s_{0}$ is introduced as the boundary separating the ground state contribution from the continuum. The spectral distributions $\rho^{(\prime)}_{1(2)} (s)$ are defined as:
		   	  \begin{equation}  
	   \rho^{(\prime)}_{1(2)} (s)=\frac{1}{\pi} \mathrm{Im} [\tilde\Pi^{(\prime)} _{1(2)} (q)].
	   	   \end{equation}
The explicit expressions for $  \rho^{(\prime)}_{1(2)} (s)$ are detailed in the Appendix for the calculations of the perturbative part and nonperturbative part of the mass dimension three, four and five for the $\Xi_{ttu}$, $\Xi_{ttd}$ and $\Omega_{tts}$ and the perturbative part and nonperturbative part of the mass dimension four for the $\Omega_{ttc}$ and $\Omega_{ttb}$ baryons. In Eqs.~(\ref{eqQCD4}) and (\ref{eqQCD5}), $\Gamma^{(\prime)}_{1(2)} (M^2)  $  are functions describing the evaluations of the nonperturbative part of the mass dimension six, seven and  eight for the   $\Xi_{ttu}$, $\Xi_{ttd}$ and $\Omega_{tts}$ and the mass dimension six and  eight  for the $\Omega_{ttc}$ and $\Omega_{ttb}$ baryons. 
Although vacuum condensates up to dimension six are explicitly listed among the input parameters, the contributions of higher-dimensional nonperturbative operators (up to dimension eight) are effectively taken into account through the standard QCD factorization hypothesis, whereby they are expressed in terms of products of lower-dimensional condensates.  The explicit expressions for the components of  $ \Gamma^{(\prime)}_{1(2)} (M^2)  $ are very lengthy and we prefer not to present them here. 
The outcomes obtained from the phenomenological and QCD representations are matched using dispersion techniques by equating the coefficients corresponding to 
the same Lorentz structures. This matching procedure yields the QCD sum rule expressions that determine the mass and the residue, which are given as follows:

\begin{eqnarray}
	\lambda^2 e^{-\frac{m^2}{M^2} }=\tilde{\Pi}^{(\prime)} _{\not\!q}(s_0,M^2),
	\label{Eq:cor1}
\end{eqnarray}
and
\begin{eqnarray}
	\lambda^2 m e^{-\frac{m^2}{M^2}} =\tilde{\Pi}^{(\prime)} _{I}(s_0,M^2).
	\label{Eq:cor2}
\end{eqnarray}
Various techniques are available for extracting the masses   of   the ground  state of baryons. In this work, we employ a particularly robust procedure that underpins our numerical analysis. Specifically, by dividing Eq. (\ref{Eq:cor2})  by Eq. (\ref{Eq:cor1})
\begin{eqnarray}
  m  = \frac{\tilde{\Pi}^{(\prime)} _{I}(s_0,M^2)}{\tilde{\Pi}^{(\prime)} _{\not\!q}(s_0,M^2)},
	\label{Eq:cor32}
\end{eqnarray}
 the masses of these doubly topped baryons for the ground state can be determined.
We present the numerical  evaluation of the QCD sum rules for the masses of the doubly topped baryons, including $\Xi_{ttu}$, $\Xi_{ttd}$, $\Omega_{tts}$, $\Omega_{ttc}$ and $\Omega_{ttb}$  in the next section.

\section {Numerical Analysis of the baryon masses}\label{sec:three}	

Here we present the numerical analysis of our results for the masses   of doubly topped baryons. The calculations make use of the input parameters summarized in Table~\ref{tab:Parameter}.  As listed in this table, vacuum condensates up to dimension six are explicitly provided; however, contributions from operators with dimensions extending to eight are taken into account in the analysis. Higher dimensional nonperturbative terms   are reduced to combinations of lower dimensional condensates by employing the standard QCD factorization assumption as previously mentioned. This treatment is applied in the construction of both heavy  and light quark propagators discussed earlier. By combining the propagators according to Eqs. (\ref{shekari}) and (\ref{eqQCD1}), the contributions of higher dimensional operators appear as products of lower dimensional operators, with their numerical values determined from the parameters provided in Table \ref{tab:Parameter}.
\begin{table}[tbp]
	\begin{tabular}{|c|c|}
		\hline
		Parameters & Values \\ \hline\hline
			$m_{t}$                                     & $172.56\pm 0.31~\mathrm{GeV}$ \cite{ParticleDataGroup:2022pth}\\ 
		$m_{b}$                                     & $4.78\pm 0.06~\mathrm{GeV}$ \cite{ParticleDataGroup:2022pth}\\
			$m_{c}$                                     & $1.67\pm 0.07~\mathrm{GeV}$ \cite{ParticleDataGroup:2022pth}\\
		$m_{s}$                                   & $93.5 \pm 0.8~\mathrm{MeV}$ \cite{ParticleDataGroup:2022pth}\\
						$m_{d}$                                   & $4.70\pm 0.07~\mathrm{MeV}$ \cite{ParticleDataGroup:2022pth}\\
							$m_{u}$                                   & $2.16\pm 0.07~\mathrm{MeV}$ \cite{ParticleDataGroup:2022pth}\\
		$\langle \bar{q}q \rangle $    & $(-0.24\pm 0.01)^3$ $\mathrm{GeV}^3$ \cite{Belyaev:1982sa}  \\
		$\langle \bar{s}s \rangle $               & $0.8\langle \bar{q}q \rangle$ \cite{Belyaev:1982sa} \\
		$m_{0}^2 $                                & $(0.8\pm0.1)$ $\mathrm{GeV}^2$ \cite{Belyaev:1982sa}\\
		$\langle \overline{q}g_s\sigma Gq\rangle$ & $m_{0}^2\langle \bar{q}q \rangle$ \cite{Belyaev:1982sa}\\
		$\langle g_s^2 G^2 \rangle $              & $4\pi^2 (0.012\pm0.004)$ $~\mathrm{GeV}
		^4 $\cite{Belyaev:1982cd}\\
		$\langle g_s^3 G^3 \rangle $                & $ (0.57\pm0.29)$ $~\mathrm{GeV}^6 $\cite{Narison:2015nxh}\\
		
		\hline\hline
	\end{tabular}%
	\caption{Parameters used in our evaluations.
	}
	\label{tab:Parameter}
\end{table}

In addition to the input quantities summarized in Table~\ref{tab:Parameter}, the QCD sum rule analysis involves three auxiliary parameters that must be specified: the mixing parameter $\beta$, the Borel mass $M^2$ and the continuum threshold $s_0$. Their working ranges are determined by imposing the conventional reliability criteria of the method, namely stability with respect to unphysical parameters, sufficient suppression of continuum and excited state effects, and a well behaved OPE.

The admissible range of the parameter $\beta$ is identified through a detailed analysis employing a parametric representation of the results as functions of $ \cos\theta$, where  $\beta = \tan\theta $. 
The analysis identifies the stability zones—those regions where the results vary minimally with respect to changes in $ \cos\theta$—as  
\begin{eqnarray}
	-1\leq\cos\theta\leq -0.5 ~~~~~\mbox{and} ~~~~~~0.5\leq \cos\theta\leq 1,
\end{eqnarray}
 for all doubly topped baryon configurations.  

\begin{figure}[h!]
	\begin{center}
		\includegraphics[totalheight=6.5cm,width=8.5cm]{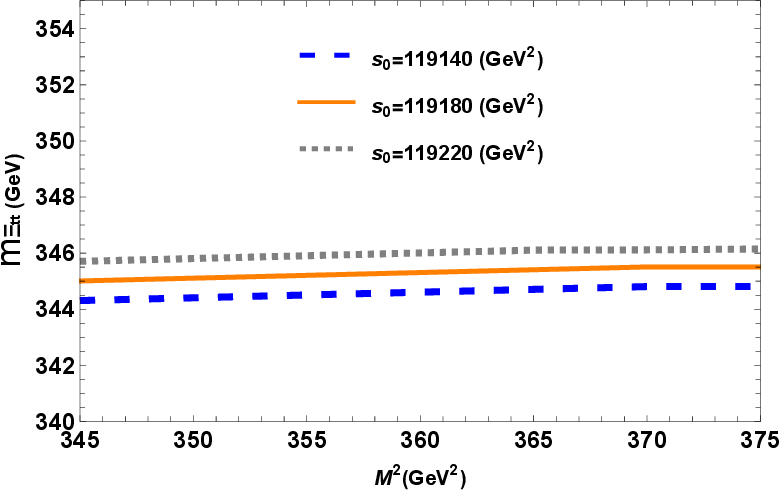}
		\includegraphics[totalheight=6.5cm,width=8.5cm]{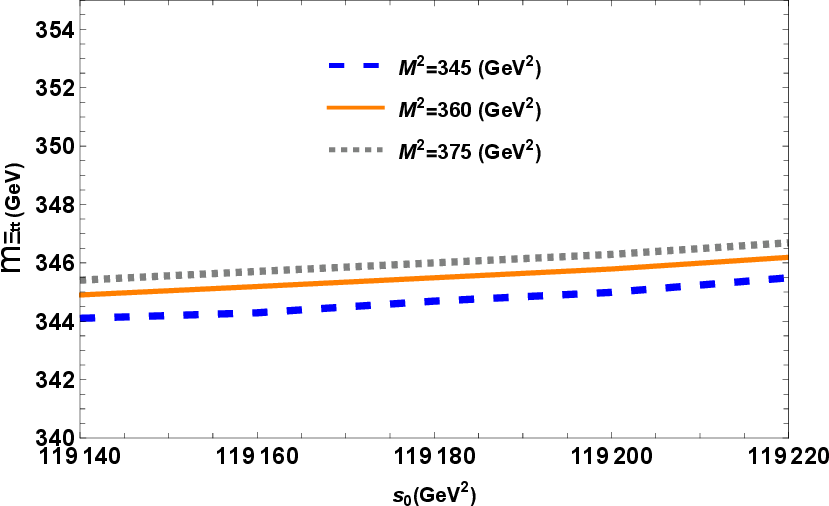}
	\end{center}
	\caption{Left: Behavior  of the extracted $\Xi_{ttu}$ mass as a function of the Borel parameter $M^2$ at several selected values of the continuum threshold $s_0$ and at Ioffe point $\cos\theta =-0.71$.
		Right: Behavior of the $\Xi_{ttu}$ mass with respect to the continuum threshold $s_0$ for different fixed choices of the Borel mass parameter $M^2$  and  at Ioffe point  $\cos\theta =-0.71$.}
	\label{1}
\end{figure}

Another requirement of the method is satisfied when the perturbative term provides the leading contribution and nonperturbative corrections decrease rapidly with increasing operator dimension. The dominance of the lowest-lying hadronic state further constrains the analysis, ensuring that the pole contribution outweighs those from higher resonances and the continuum. As a result, the maximum value of  $M^2$ is fixed by demanding that the pole contribution remain larger than the combined continuum and excited state contributions. Under these considerations, the following conditions are imposed:
 \begin{eqnarray}
	\mathrm{PC}=\frac{\tilde{\Pi}^{(\prime)}(s_0,M^2)}{\tilde{\Pi}^{(\prime)}(\infty,M^2)}\geq \frac{1}{2}.
\end{eqnarray}
The minimum value of the Borel parameter  $M^2$ is fixed by requiring a convergent operator product expansion. This condition implies that the perturbative part must provide the leading contribution, while the effects of nonperturbative condensates become progressively suppressed as their operator dimension increases. To quantify this constraint, we introduce the following ratio:
\begin{equation}
	\frac{\tilde{\Pi} ^{\mathrm{(\prime),Dim8}}(s_0,M^2)}{\tilde{\Pi} ^{(\prime)}(s_0,M^2)}\le\ 0.05.
	\label{eq:Convergence}
\end{equation}  	
The continuum threshold $s_0$, associated with the emergence of excited and continuum contributions, is chosen such that the lowest-lying state provides the dominant contribution in the sum rule analysis. This choice effectively suppresses the influence of higher resonances and the continuum. The working windows for the Borel parameter  and the threshold  in each channel are summarized in Table~\ref{tab:res}. Our numerical study shows that the extracted observables exhibit only a weak sensitivity to variations of these auxiliary parameters within their allowed ranges. 
To visualize this behavior, as an example, the dependence of the mass on $M^2$ and $s_0$   is displayed in Fig. \ref{1}  for  the   $\Xi_{ttu}$ baryon. The results illustrate a high level of stability throughout the selected intervals of the Borel mass and continuum threshold.  

After fixing suitable ranges for the auxiliary parameters, we report the resulting mass predictions for the $\Xi_{ttu}$, $\Xi_{ttd}$, $\Omega_{tts}$, $\Omega_{ttc}$ and $\Omega_{ttb}$ baryons in Table \ref{tab:res}. The associated errors are evaluated by systematically accounting for uncertainties inherent to the sum rule analysis, including the quark–hadron duality approximation, the dependence on the Borel mass $M^{2}$ and the continuum threshold $s_{0}$, together with the propagated uncertainties of the input quantities, such as the   quark mass and the relevant nonperturbative condensates.

\begin{table}[ht!]
\centering
\renewcommand{\arraystretch}{0.85} 
\setlength{\tabcolsep}{6pt}       
\fontsize{9}{11}\selectfont
\begin{tabular}{|c|c|c|c|}
\hline
Baryon& $M^2~(\mathrm{GeV^2})$&$s_0(\mathrm{GeV^2})$  & m $(\mathrm{GeV})$ \\
\hline \hline
$\Xi_{ttu}$ & 345-375 & 119140-119220 & $345.20{}^{+1.10}_{-1.11}$ \\ \hline
$\Xi_{ttd}$ & 345-375 & 119140-119220 & $345.20{}^{+1.10}_{-1.11}$ \\ \hline
$\Omega_{tts}$ &345-375&  119150-119230 & $345.22{}^{+1.11}_{-1.13}$ \\ 
\hline 
$\Omega_{ttc}$ &347-377& 120300-120400& $346.88{}^{+1.19}_{-1.17}$ \\ 
\hline 
$\Omega_{ttb}$ &350-380& 122300-122400&  $350.00{}^{+1.21}_{-1.18}$\\
\hline \hline 
\end{tabular}
\caption{ Mass of the doubly topped baryons using QCD sum rules, specifically within the  established Borel and continuum stability criteria.}
\label{tab:res}
\end{table}
The sums of the constituent-quark masses for the baryons $\Xi_{ttu}$, $\Xi_{ttd}$, $\Omega_{tts}$, $\Omega_{ttc}$, and $\Omega_{ttb}$ are found to be $345.12\pm0.62~\mathrm{GeV}$, $345.12\pm0.62~\mathrm{GeV}$, $345.21\pm0.62~\mathrm{GeV}$, $346.79\pm0.69~\mathrm{GeV}$, and $349.82\pm0.70~\mathrm{GeV}$, respectively. As summarized in Table~\ref{tab:res}, the extracted masses of all considered baryonic states slightly exceed the corresponding sums of their constituent quark masses. Nevertheless, owing to the associated uncertainties, the lower bounds of the predicted mass intervals extend below these reference values. This behavior should not be interpreted as evidence for a physically negative binding energy or the existence of deeply bound states. Rather, it reflects the systematic uncertainties intrinsic to the QCD sum rule framework, including sensitivities to the Borel mass parameter and the continuum threshold. Such effects can naturally lead to modest fluctuations around the naive constituent mass sums without implying any physical binding scenario. From a physical standpoint, these baryonic systems are constrained to form color-singlet configurations and are stabilized by strong multi-quark interactions. It is therefore plausible that nontrivial quantum correlations or collective effects among the three quarks play a role in maintaining the color-singlet structure and ensuring the dynamical stability of the baryons. A quantitative assessment of the relevance of such mechanisms, however, requires more refined theoretical analyses and, ultimately, dedicated experimental input.

\section {Conclusion}\label{sec:four}		
The top quark, characterized by its exceptionally large mass and ultrashort lifetime, represents a unique sector of the SM. While its rapid decay has traditionally been viewed as prohibitive for the formation of bound states, this same interplay between mass and lifetime renders the top quark an especially sensitive probe of short-distance QCD dynamics and potential physics beyond the SM. As a result, the top-quark sector continues to provide a compelling framework for testing fundamental interactions at the highest accessible energy scales.

	Following the experimental discovery of the doubly charmed baryon $ \Xi^{++}_{cc} (3621)$   by the LHCb collaboration in 2017, extensive experimental efforts have been devoted to the search for the remaining members of this baryonic family with two heavy quarks. Motivated by recent advances in the study of   $t\bar{t}$ 	system observed by the CMS and ATLAS collaborations, we conjecture—as a theoretical extrapolation given the extremely short lifetime of the top quark—that baryons containing two heavy top quarks could, in principle, also be realized.

This research presents a comprehensive theoretical determination of the masses for baryonic states  incorporating two heavy top quarks, leveraging the  QCD  sum rule methodology. We specifically focused on five distinct systems: $\Xi_{ttu}$, $\Xi_{ttd}$, $\Omega_{tts}$, $\Omega_{ttc}$ and $\Omega_{ttb}$ baryons. Based on our results, the extracted central masses are slightly above the sum of the constituent quark masses, consistent with the inherent uncertainties of the method. These systems are required to form color singlet configurations and are bound by the strong interaction through  multi-quark forces. As said,  although the central values of the predicted masses lie slightly above the corresponding constituent quark masses thresholds, the associated uncertainty bands extend below these reference values. This behavior should be understood as a manifestation of the intrinsic systematic uncertainties of the QCD sum rule approach, most notably the sensitivity to the Borel mass parameter and the continuum threshold,  \textit{rather than as an indication of a physical  binding mechanism}.
From a conceptual standpoint, it is conceivable that nontrivial quantum correlations or collective entanglement effects among the three quarks play a role in maintaining the overall color singlet structure and contributing to the dynamical stability of the baryonic state. A definitive assessment of the significance of such effects, however, necessitates more sophisticated theoretical frameworks as well as dedicated experimental investigations.

In addition to calculating the mass predictions, the findings of this study offer   structural insights regarding the potential internal quantum dynamics of these exceptionally configurations containing two top quarks.
Ultimately, the predicted masses—especially the first quantitative estimates for $\Xi_{ttu}$, $\Xi_{ttd}$, $\Omega_{tts}$, $\Omega_{ttc}$ and $\Omega_{ttb}$ —form essential theoretical benchmarks that can guide future high-energy collider experiments seeking to identify and characterize such heavy baryons.

\section*{Consent to Participate declaration} 
Consent to Participate declaration: not applicable.

\section*{Consent to Publish declaration} 
Consent to Publish declaration: not applicable.

\section*{Author Contribution declaration}
Author Contribution declaration: both authors have equal contributions in defining the problem, literature search, solving the problem, numerical analyses,  interpretation of the results and preparation of the manuscript.

\section*{Ethics declaration}
Ethics declaration: not applicable.

\section*{Data Availability Statement} 
Data Availability Statement: No Data associated in the manuscript. We do not use or produce any data.  All the information are included in the text,  formulas, tables and figures. 

\section*{Competing Interest declaration} 
 There is no "Competing Interest" regarding this manuscript.

\section*{Funding Declaration} 
K. Azizi thanks Iran national science foundation (INSF) for the partial financial support provided under the elites Grant No. 40405095.

\section*{APPENDIX: QCD side expressions of the calculations}

In this appendix, we present the expressions of the spectral densities obtained from the calculations.  $Q$	illustrates  top quark field,  $Q'$ indicates bottom or charm quark field and $q$ represents the light quark fields. The resulting expressions for the QCD side are ultimately written in the following form for the   $\Xi_{ttu}$, $\Xi_{ttd}$ and $\Omega_{tts}$ baryons:

\begin{align}
	\rho^{(pert)}_{1}(s)&=\frac{1}{128 \pi^4}	\int_{0}^{1} \, du \int_{0}^{1-u}
	dv \,  \frac {-3  \,  D_1  \,  \Theta(D_1) } {Z_1^4 \, Z_3^2} \nonumber\\
	&\Bigg\{ \Bigg(-2 \, m_{Q} \, u \, Z_1^2 \, (m_{Q'} \, v - 3 \,  m_q \, Z_2) \, Z_3 -  v \, Z_2 \, \Big(-6 \, m_{Q'} \, m_q \, Z_1^2 \, Z_3 + 5 \, u \, (3 \, D_1 \, Z_1^2 + 2 \, s  \, u  \, v \, Z_2 \, Z_3) \Big) \Bigg)  \nonumber\\
	&+2 \, \beta \, u \,  v \,  \Bigg(-3 \, D_1 \, Z_1^2 \, Z_2 - 2 (-m_{Q} \, m_{Q'} \, Z_1^2 + s \, u \, v \, Z_2^2) Z_3 \Bigg)\nonumber\\
	&-  \beta^2 \Bigg(-2 \, m_{Q} \, u \, Z_1^2 \, (-m_{Q'} \, v - 3 \, m_q \, Z_2) \, Z_3 - 
	v \, Z_2 \, \Big(-6 \, m_{Q'} \, m_q \, Z_1^2 \,  Z_3 - 5 \, u (3 \, D_1 \, Z_1^2 + 2 \, s \, u \, v \, Z_2 \, Z_3) \Big) \Bigg) \Bigg\},
\end{align}

\begin{align}
	\rho^{(pert)}_{2}(s)&=\frac{1}{64 \pi^4}	\int_{0}^{1} \, du \int_{0}^{1-u}
	dv \,  \frac {3  \,  D_1  \,  \Theta(D_1) } {Z_1^3 \, Z_3^2} \nonumber\\
	&\Bigg\{ \Bigg( D_1 \, Z_1^2 \, \Big(3 (m_{Q} \, u + m_{Q'} \, v) - m_q \, Z_2 \Big) - \Big(-s \, u \, v \, Z_2  \, (3 \, m_{Q}  \, u - m_q \, Z_2) + 
	m_{Q'} \, (5 \, m_{Q} \, m_q \, Z_1^2 - 3 \, s \, u \, v^2 \, Z_2) \Big) \, Z_3 \Bigg) \nonumber\\
	&-2 \, \beta \, m_q \Bigg(-D_1 \, Z_1^2 \,  Z_2 - (-m_{Q'} \, m_{Q} \, Z_1^2 + s \, u \, v \, Z_2^2) \, Z_3 \Bigg)\nonumber\\
	&+ \beta^2   \, \Bigg(D_1 \, Z_1^2 \, \Big(-3 \, (m_{Q} \, u + m_{Q'} \,  v) - m_q  \, Z_2 \Big) - \Big(-s \, u \, v  \,Z_2 \, (-3 \, m_{Q} \, u - m_q \, Z_2) + m_{Q'} \, (5 \, m_{Q} \, m_q \, Z_1^2 \nonumber\\
	&+ 3 \, s \, u \, v^2 \, Z_2) \Big) \, Z_3 \Bigg) \Bigg\},
\end{align}

\begin{align}
	\rho^{(3d)}_{1}(s)&=\frac{3}{32 \pi^4}	\int_{0}^{1} \, du \int_{0}^{1-u}
	dv \,    \,  \langle \overline{q} q \rangle \,  \Theta(D_2) \Bigg\{ 2 \, m_{Q} \, u + (2 \, m_{Q'}  + 5 \, m_q \, u) \, Z_4 +2 \, \beta \, ( m_q  u Z_4) + \beta^2\, \Big(-2  \, m_{Q}   \, u \nonumber\\
	&- (2  \, m_{Q'}   -5 \, m_q  \, u )Z_4 \Big) \Bigg\},&&&&&&&&&
\end{align}

\begin{align}
	\rho^{(3d)}_{2}(s)&=\frac{1}{16 \pi^4}	\int_{0}^{1} \, du \int_{0}^{1-u}
	dv \, \langle \overline{q} q \rangle \,  \Theta(D_2)  \Bigg\{  - \Big(2  \, D_2 + 5  \, m_{Q'}   \, m_{Q}  + u  \, (3 m_{Q}   \, m_q + s - s \,  u) + 3  \, m_{Q'}   \, m_q  \, Z_4  \Big) \nonumber\\
	&+2 \, \beta \, (2   \, D_2 - m_{Q'}   \, m_{Q} + s  \,  u  \, Z_4) + \beta^2\, \Big(-2   \, D_2 - 5   \, m_{Q'}   \, m_{Q} + 3   \, m_{Q}   \, m_q   \, u + (3 m_{Q'}   \, m_q  - s  \,  u )Z_4 \Big) \Bigg\},&&&&&&&&&
\end{align}

\begin{align}
	\rho^{(4d)}_{1}(s)&=\frac{1}{256 \pi^4}	\int_{0}^{1}  \, du \int_{0}^{1-u}
	dv \,  \Big\langle\frac{\alpha_{s}GG}{\pi}\Big\rangle\ \,  \frac {-3 \, u \, v\,  Z_2   \,  \Theta(D_3) } {Z_1^4}\nonumber\\
	&\Bigg\{\Big(3 u^2 + u (-3 + 7 \, v) - 3 \,  v  \, Z_3 \Big)+ 2 \, \beta  \Big(u^2 + u (-1 + 3 \, v) - v \, Z_3\Big)
	+\beta^2 \, \Big(3 u^2 + u (-3 + 7 \, v) - 3 \, v \, Z_3 \Big) \Bigg\},&&&&&&&&&&&&&&&&&&&&&&&&&&&&&&&&&&&&&&&&&&&&&&&&&&&&&
\end{align}

\begin{align}
	\rho^{(4d)}_{2}(s)&=\frac{1}{128 \pi^4}	\int_{0}^{1}  \, du \int_{0}^{1-u}
	dv \,  \Big\langle\frac{\alpha_{s}GG}{\pi}\Big\rangle\ \,  \frac {    \Theta(D_3) } {Z_1^4}\nonumber\\
	&\Bigg\{ -\Bigg[m_{Q}  \,  u  \, Z_1 \Big(u^2 + u (-1 + 5 \,  v) - 3  \, v  \, Z_3\Big) + 
	v \Big[\Big(m_q  \, u + m_{Q'}  \,  (2 + 8 \,  u) \Big) v^3 - 
	3 \, m_{Q'}  \,  v^4 \nonumber\\
	&- \Bigg( (4  \, m_{Q'}  \,  + m_q) u (-1 + 2 \,  u) v  + \Big(2  \, m_q  \, u + 
	m_{Q'}  \,  (-1 + 11  \, u) \Big) v^2 \Bigg)Z_4 + (3 \,  m_{Q'}  \,  + m_q) u^2  \, Z_4^2 \Big] \Bigg]\nonumber\\
	&- 2 \, \beta  \Big(m_q \, u \, v  \, Z_1 \, Z_2 \Big)
	+\beta^2 \, \Bigg[m_{Q} u Z_1 \Big(u^2 + u (-1 + 5 v) - 3 v Z_3 \Big) + 
	v \Bigg(m_q u Z_1 Z_2 + 
	m_{Q'} \Big(2 (1 + 4 \,  u) v^3 \nonumber\\
	&- 3  \, v^4 - \Big(
	4  \, u (-1 + 2  \, u) v   + (-1 + 11  \, u) v^2  \Big)\, Z_4 + 3  \, u^2  \, Z_4^2 \Big)\Bigg) \Bigg] \Bigg\},&&&&&&&&&&&&&&&&&&&&&&&&&&&&&&&&&&&&
\end{align}

\begin{align}
	\rho^{(5d)}_{1}(s)=0,&&&&&&&&&&&&&&&&&&&&&&&&&&&&&&&&&&&&&&&&&&&&&&&&&&&&&&&&&&&&&&&&&&&&&&&&&&&&&&&&&&&&&&&&&&&&&&&&&&&&&&&&&&&&&&&&&&&&&&&&&&
\end{align}

\begin{align}
	\rho^{(5d)}_{2}(s)&=\frac{5}{64 \pi^4}	\int_{0}^{1} \, du \int_{0}^{1-u}
	dv \,   \, \langle \overline{q}g_s\sigma Gq\rangle \, u \, Z_4   \,  \Theta(D_4)  \Big(1-2 \, \beta+ \beta^2 \Big),&&&&&&&&&&&&&&&&&&&&&&&&&&&&&&&&&&&&&&&&&&&&&&&&&&&&&&&&&&&&
\end{align}

where $  \Theta(...) $ is the Unit-Step function and

\begin{eqnarray}
	D_1&=&-\frac{Z_3}{Z_1^2} \Big((m_{Q}^2 \, u + m_{Q'}^2 \, v)Z_1 + s \, u \, v \, Z_2 \Big),\nonumber\\
	D_2&=&-m_{Q}^2 u - (m_{Q'}^2  - s u )Z_4,\nonumber\\
	D_3&=&\frac{Z_3}{Z_1^2} \Big((m_{Q}^2 \, u  + m_{Q'}^2 \, v) Z_1 + s \, u \, v \, Z_2 \Big) ,\nonumber\\
	D_4&=&-m_{Q}^2 \,  u - (m_{Q'}^2  - s \,  u )Z_4,
\end{eqnarray}

and we have defined

\begin{eqnarray}
	Z_1&=& u^2 + u (-1 + v) + (-1 + v) \, v,\nonumber\\
	Z_2&=&1 - u - v,\nonumber\\
	Z_3&=&1 - v,\nonumber\\
	Z_4&=&1 -u.
\end{eqnarray}

Here, for the $\Omega_{ttc}$ and $\Omega_{ttb}$ baryons we define the spectral densities in following forms:	   
\begin{equation}
	\begin{split}
		\rho'^{(pert)}_{1}(s)&=\frac{3}{64 \pi ^4}
		\int^{1}_{0} dz \int^{1-z}_{0} dr \ \frac{H\,\Theta(H)}{P_1^4\, P_3^3} \Bigg\{\bigg[2\, m_Q^2\, P_1^2\, P_2\, P_3^2\, z + 2 \,m_Q\, m_{Q'}\, P_1^2 \,P_3 \,(P_3 + 4\, z) \,(P_1 + P_2\, z) \\
		&+ 
		z \,(10\, s\, P_2\, (P_3 - z) \,z \,(P_1 + P_2\, z)^2 - 
		H\, P_1^2\, (P_1\, (P_2 + 14\, P_3 - 14\, z) + 15\, P_2\, (P_3 - z)\, z))\bigg]\\
		&+2 \beta\bigg[-2\, m_Q^2\, P_1^2\, P_2\, P_3^2 + 2\, s\, P_2 \,(P_3 - z) \,z\, (P_1 + P_2\, z)^2 + 
		H\, P_1^2\, (3 \,P_2\, z \,(-P_3 + z)\\
		& + P_1\, (P_2 - 4\, P_3 + 4\, z))\bigg]	+	\beta^2\bigg[-2\, m_Q^2\, P_1^2\, P_2\, P_3^2\, z + 2\, m_Q \,m_{Q'}\, P_1^2\, P_3\, (P_3 + 4\, z) (P_1 + P_2\, z)\\
		& + 
		z\, (-10\, s\, P_2 (P_3 - z)\, z \,(P_1 + P_2 \,z)^2 + 
		H \,P_1^2 (P_1\, (P_2 + 14\, P_3 - 14\, z) + 15\, P_2\, (P_3 - z)\, z)) \bigg]
		\Bigg\},
	\end{split}
\end{equation}
\begin{equation}
	\begin{split}
		\rho'^{(pert)}_2(s)&=\frac{3}{32 \pi ^4}
		\int^{1}_{0} dz \int^{1-z}_{0} dr \ \frac{H\, \Theta (H)}{P_1^3\, P_3^3} \Bigg\{\bigg[ -5 \,m_Q^2\, m_{Q'}\, P_1^2\, P_3^2 - m_Q \,s \,P_2 \,P_3\, (5 \,P_3 - 4\, z)\, z\, (P_1 + P_2 \,z) \\
		&- 
		m_{Q'}\, s\, (P_3 - z) \,z \,(P_1 + P_2\, z)^2 + 
		H \,P_1^2\, (m_Q \,P_3 \,(5\, P_3 - 4\, z) + m_{Q'}\, (P_1 + (P_3 - z)\, z))
		\bigg]\\
		&-2 \beta\,m_{Q'} \bigg[ m_Q^2\, P_1^2 \,P_3^2 - s\, (P_3 - z) \,z\, (P_1 + P_2\, z)^2 + 
		H\, P_1^2\, (P_1 + (P_3 - z)\, z)\bigg]\\
		&+
		\beta^2\bigg[-5\, m_Q^2\, m_{Q'}\, P_1^2\, P_3^2 + m_Q \,s\, P_2\, P_3\, (5\, P_3 - 4\, z)\, z \,(P_1 + P_2\, z)\\
		& - 
		m_{Q'}\, s\, (P_3 - z) \,z\, (P_1 + P_2\, z)^2 + 
		H\, P_1^2\, (m_Q\, P_3\, (-5\, P_3 + 4\, z)\\
		& + m_{Q'} (P_1 + (P_3 - z)\, z))\bigg]
		\Bigg\},
	\end{split}
\end{equation}
\begin{equation}
	\begin{split}
		\rho'^{(4d)}_{1}(s)&=\frac{1}{192 \pi ^2}
		\int^{1}_{0} dz \int^{1-z}_{0} dr \ \Big\langle\frac{\alpha_{s}GG}{\pi}\Big\rangle\ \frac{\Theta(H)}{P_1^4\, P_3} \Bigg\{\bigg[8\,z\Big(-9 \,r^2\, (P_3 - z)\, (P_1 + P_2 \,z) + 
		3\, P_4\, z\, (3\, P_2\, z\, (-P_3 + z)\\
		& + P_1 \,(P_2 - 2\, P_3 + 2 \,z)) + 
		r \,(P_1\, P_3\, (9 + 68 \,z) - P_1 \,z\, (9 + 9 \,P_2 + 68\, z) + 
		P_2\, (P_3 - z) \,z \,(9 + 77\, z))\Big)\bigg]\\
		&+2\, \beta \bigg[3\,z\Big(P_4 \,z\, (27\, P_1\, P_2 + 50\, P_1\, (-P_3 + z) + 77 \,P_2\, z\, (-P_3 + z)) + 
		11\, r^2 \,(7\, P_2\, z\, (-P_3 + z)\\
		& + P_1\, (P_2 - 6\, P_3 + 6\, z)) + 
		r\, (-11\, P_1\, P_2 + 6\, P_1\, (P_3 - z) (11 + 12\, z) + P_2\, (P_3 - z)\, z \,(77 + 72\, z))\Big)\bigg]\\
		&	+	\beta^2\bigg[8\,z\Big(-9\, r^2\, (P_3 - z)\, (P_1 + P_2\, z) + 
		3\, P_4 \,z\, (3\, P_2\, z\, (-P_3 + z) + P_1 \,(P_2 - 2\, P_3 + 2\, z)) \\
		&+ 
		r\, (P_1 \,P_3\, (9 + 68\, z) - P_1\, z\, (9 + 9\, P_2 + 68 \,z) + 
		P_2\, (P_3 - z)\, z \,(9 + 77\, z))\Big) \bigg]
		\Bigg\},
	\end{split}
\end{equation}
\begin{equation}
	\begin{split}
		\rho'^{(4d)}_2(s)&=\frac{1}{192 \pi ^2}
		\int^{1}_{0} dz \int^{1-z}_{0} dr \ \Big\langle\frac{\alpha_{s}GG}{\pi}\Big\rangle\ \frac{\Theta(H)}{P_1^4\, P_3} \Bigg\{\bigg[m_Q \Big(6\, (P_3 - z)\, z^2\, (-3\, P_3^3 + 9\, P_3^2\, z - (9 - 9\, r + P_3)\, z^2 + 4\, z^3)\\
		& +
		P_1\, (-3\, P_3^3 \,(r + 4\, P_3) + 
		P_3^2\, (3 - 31\, r + 54\, P_3)\, z + (-36 + 64\, r - 57\, P_3)\, P_3\, z^2 + 
		68\, P_3\, z^3 - 24 \,z^4)\Big)\\
		& + 
		m_{Q'}\Big(-3\, r^2\, P_1 (P_1 + z\, (-P_3 + z)) - 3 \,P_1\, P_4\, z\, (P_1 + z\, (-P_3 + z)) + 
		3\, r\, P_1 \,(1 + z) \,(P_1 + z\, (-P_3 + z)) \\
		&+ 
		2\, r^3\, P_3\, (2\, P_1 + 3\, z\, (-P_3 + z))\Big)\bigg]
		+2\, \beta \bigg[m_{Q'}\,\Big(-3\, r^2\, P_1\, (P_1 + z\, (-P_3 + z)) - 3\, P_1\, P_4\, z\, (P_1 + z\, (-P_3 + z)) \\
		&+ 
		3\, r\, P_1 (1 + z) (P_1 + z\, (-P_3 + z)) + 2\, r^3 \,P_3\, (2\, P_1 + 3 \,z (-P_3 + z))\Big)\bigg]\\
		&	+	\beta^2\bigg[m_Q\,\Big(6\, (P_3 - z) \,z^2\, (5\, P_3^3 - 15\, P_3^2\, z + (15 - 15\, r + P_3)\, z^2 - 
		6 \,z^3) + 
		P_1\, (-3\, (-1 + r)\, r\, P_3^2 + 20\, P_3^4\\
		& + 
		P_3^2\, (31\, r - 3\, (30\, P_3 + P_4))\, z + 
		2\, P_3\, (30 - 44\, r + 45\, P_3)\, z^2 - 112 \,P_3\, z^3 + 36\, z^4)\Big)\\
		& + 
		m_{Q'}\, \Big(-3 \,r^2 \,P_1\, (P_1 + z\, (-P_3 + z)) - 3\, P_1\, P_4\, z \,(P_1 + z\, (-P_3 + z)) + 
		3 \,r\, P_1 (1 + z)\, (P_1 + z (-P_3 + z))\\
		& + 
		2 \,r^3\, P_3 \,(2\, P_1 + 3\, z\, (-P_3 + z))\Big) \bigg]
		\Bigg\},
	\end{split}
\end{equation}

where 
\begin{eqnarray}
	H &=&\frac{P_3}{P_1^2} \Big((m_{Q}^2 \, P_3 + m_{Q'}^2 \, r)\,P_1 + s \,r\,z\,P_3 \Big),\nonumber\\
	P_1&=& r^2 + r\, (-1 + z) + (-1 + z) \, z,\nonumber\\
	P_2&=&1 -r - z,\nonumber\\
	P_3&=&1 -r,\nonumber\\
	P_4&=&1 -z.
\end{eqnarray}
%


\end{document}